\documentclass[11pt]{article}
\usepackage{amssymb}
\usepackage{multirow}
\usepackage{amsmath}
\usepackage{amsfonts}
 \usepackage{pifont}
\usepackage{amsmath,amsfonts,amssymb,theorem,graphicx,listings,txfonts}
\usepackage{graphics}
 \usepackage{caption2}
\usepackage{flafter}
 \usepackage{indentfirst}
\usepackage{epsfig}
  \usepackage{mathrsfs}
  \usepackage{float}
\usepackage{subfigure}
 \usepackage{cite}
\newtheorem{theorem}{Theorem}[section]

\newtheorem{proposition}[theorem]{Proposition}

\newtheorem{definition}[theorem]{Definition}

\theoremstyle{example}
\newtheorem{example}[theorem]{Example}
\theoremstyle{programme}

\theoremstyle{property}

\theoremstyle{problem}

\renewcommand{\arraystretch}{1}

\title{Comments on ``Characteristic matrix of
covering and its application to Boolean matrix decomposition[Information Sciences 263(1),
186-197, 2014]"}

\author
{Guangming Lang$^{1,2,3}$
\thanks{Corresponding author.\quad 
\newline\mbox{}\hspace{0.55cm}
E-mail address: langguangming1984@tongji.edu.cn(G.M.Lang). }\hspace{1cm}
\\
\small {$^{1}$ Department of Computer Science and Technology, Tongji University}\\
\small {Shanghai 201804, P.R. China}\\
\small {$^{2}$ School of Mathematics and Statistics, Changsha University of Science and Technology}\\
\small {Changsha, Hunan 410114, P.R. China}\\
\small {$^{3}$  The Key Laboratory of Embedded System and Service Computing, Ministry of Education, Tongji University}\\
\small {Shanghai 201804, P.R. China}\\
}
\date{}

\begin{document}
\maketitle \baselineskip=17pt
\begin{center}
\begin{quote}
{{\bf Abstract.}
In this note, we show some improvements for	
Theorem 7 and Example 8 in Shiping Wang[Information Sciences 263(1),
186-197, 2014]. Concretely, we study further the sixth lower and upper approximations of sets for covering approximation spaces. Furthermore, we present the sixth dual lower and upper approximations of sets for covering approximation spaces. We also construct the sixth dual lower and upper approximations of sets from the view of matrix. Throughout, we 	
use the same notations as Shiping Wang[Information Sciences 263(1),
186-197, 2014].

\\}
\end{quote}
\end{center}
\renewcommand{\thesection}{\arabic{section}}

\section{Introduction}

Wang et al.\cite{Wang1} transformed
the computation of the second, fifth, and sixth lower and upper approximations of a set into products of the
characteristic matrices and characteristic function of the set, which provides an effective tool for computing the second, fifth, and sixth lower and upper approximations of sets from the view of matrix.
Especially, Wang et al.\cite{Wang1} presented Theorem 7 for the sixth lower and upper approximations of sets in covering approximation spaces. They also provided Example 8 to illustrate how to compute the sixth lower and upper approximations of sets. In this note, we show some improvements for	
Theorem 7 and Example 8 in Shiping Wang[Information Sciences 263(1),
186-197, 2014].

The rest of this note is organized as follows: Section 2 briefly
reviews the second, fifth, and sixth lower and upper approximations of sets. In
Section 3, we present Theorem 7 and Example 8 in Shiping Wang[Information Sciences 263(1),
186-197, 2014].  We also provide some improvements for Theorem 7 and Example 8. Concluding remarks are shown in Section 4.

\section{Preliminaries}

In this section, we review the second, fifth, and sixth lower and upper approximations of sets
in covering approximation spaces.

\begin{definition}\cite{Wang1}
Let $(U,\mathscr{C})$ be a covering approximation space, where $U=\{x_{1},x_{2},...,x_{n}\}$,
$\mathscr{C}=\{C_{1},C_{2},...,C_{m}\}$, and $N(x)=\bigcap\{C_{i}|x\in C_{i}\in \mathscr{C}\}$ for any $x\in U$. For any
$X\subseteq U$, the second, fifth, and sixth upper and lower approximations
of $X$ with respect to $\mathscr{C}$ are defined as follows:

$(1)$ $SH_{\mathscr{C}}(X)=\bigcup\{C\in\mathscr{C}|C\cap X\neq
\emptyset\}, SL_{\mathscr{C}}(X)=[SH_{\mathscr{C}}(X^{c})]^{c};$

$(2)$ $IH_{\mathscr{C}}(X)=\{x\in U|N(x)\cap X\neq \emptyset\},
IL_{\mathscr{C}}(X)=\{x\in U|N(x)\subseteq X\};$

$(3)$ $XH_{\mathscr{C}}(X)=\bigcup\{N(x)|N(x)\cap X\neq \emptyset\},
XL_{\mathscr{C}}(X)=\bigcup\{N(x)|N(x)\subseteq X\}.$
\end{definition}

Wang et al.\cite{Wang1} presented the concepts of the type-1 and type-2
characteristic matrices for computing the second, fifth, and sixth lower and upper approximations of sets as follows:

\begin{definition}\cite{Wang1}
Let $(U,\mathscr{C})$ be a covering approximation space, where $U=\{x_{1},x_{2},...,x_{n}\}$,
$\mathscr{C}=\{C_{1},C_{2},...,C_{m}\}$, and $M_{\mathscr{C}}=(a_{ij})_{n\times m}$, where
$a_{ij}=\left\{
\begin{array}{ccc}
1,&{\rm}& x_{i}\in C_{j};\\
0,&{\rm}& x_{i}\notin C_{j}.
\end{array}
\right.$ Then

$(1)$ $\Gamma(\mathscr{C})=M_{\mathscr{C}}\cdot
M_{\mathscr{C}}^{T}=(b_{ij})_{n\times n}$ is called the type-1
characteristic matrix of $\mathscr{C}$, where
$b_{ij}=\bigvee^{m}_{k=1}(a_{ik}\cdot a_{jk})$;

$(2)$ $\prod(\mathscr{C})=M_{\mathscr{C}}\odot
M_{\mathscr{C}}^{T}=(c_{ij})_{n\times n}$ is called the type-2
characteristic matrix of $\mathscr{C}$, where
$c_{ij}=\bigwedge^{m}_{k=1}(a_{kj}-a_{ik}+1)$.
\end{definition}

By Definition 2.2, we have the characteristic function $\mathcal {X}_{X}
=\left[\begin{array}{cccccc}
  a_{1}&a_{2}&.&.&. & a_{n} \\
  \end{array}
\right]^{T}$ for any $X\subseteq U$, where $a_{i}=\left\{
\begin{array}{ccc}
1,&{\rm}& x_{i}\in X;\\
0,&{\rm}& x_{i}\notin X.
\end{array}
\right. $

In what follows, Wang et al.\cite{Wang1} provided the descriptions of the
second, fifth, and sixth lower and upper approximation operators from the view of matrix.

\begin{theorem}\cite{Wang1}
Let $(U,\mathscr{C})$ be a covering approximation space, where $U=\{x_{1},x_{2},...,x_{n}\}$,
$\mathscr{C}=\{C_{1},C_{2},...,C_{m}\}$, and
$\mathcal {X}_{X}$ the characteristic function of $X\subseteq U$. Then

$(1)$ $\mathcal {X}_{SH_{\mathscr{C}}(X)}=\Gamma(\mathscr{C})\bullet \mathcal
{X}_{X}$, $\mathcal {X}_{SL_{\mathscr{C}}(X)}=\Gamma(\mathscr{C})\odot \mathcal
{X}_{X}$;

$(2)$ $\mathcal {X}_{IH_{\mathscr{C}}(X)}=\prod(\mathscr{C})\bullet
\mathcal {X}_{X}$, $\mathcal {X}_{IL_{\mathscr{C}}(X)}=\prod(\mathscr{C})\odot
\mathcal {X}_{X}$.
\end{theorem}

\section{Main Results}

In this section, we provide Theorem 3.1 and Example 3.2(Theorem 7 and Example 8 in Shiping Wang[Information Sciences 263(1),
186-197, 2014]) as follows.

\begin{theorem}\cite{Wang1}
Let $(U,\mathscr{C})$ be a covering approximation space, where $U=\{x_{1},x_{2},...,x_{n}\}$,
$\mathscr{C}=\{C_{1},C_{2},...,C_{m}\}$, and
$\mathcal {X}_{X}$ the characteristic function of $X\subseteq U$. Then
\begin{eqnarray*}
\mathcal {X}_{XH_{\mathscr{C}}(X)}=(\prod(\mathscr{C}))^{T}\bullet\prod(\mathscr{C})\bullet
\mathcal {X}_{X}, \mathcal {X}_{XL_{\mathscr{C}}(X)}=(\prod(\mathscr{C}))^{T}\bullet\prod(\mathscr{C})\odot
\mathcal {X}_{X}.
\end{eqnarray*}
\end{theorem}
{\bf{Proof:}} For a covering $\mathscr{C}$ of $U$, we construct a special covering
$Cov(\mathscr{C})=\{N(x)|x\in U\}$ induced by $\mathscr{C}$. Then $XH_{\mathscr{C}}(X)=SH_{Cov(\mathscr{C})}(X)$ and $XL_{\mathscr{C}}(X)=SL_{Cov(\mathscr{C})}(X)$ for all $X\subseteq U$. Because $(\prod(\mathscr{C}))^{T}$ is a matrix representing $Cov(\mathscr{C})$, according to Theorem 2.3, $\mathcal {X}_{XH_{\mathscr{C}}(X)}=(\prod(\mathscr{C}))^{T}\bullet\prod(\mathscr{C})\bullet
\mathcal {X}_{X}, \mathcal {X}_{XL_{\mathscr{C}}(X)}=(\prod(\mathscr{C}))^{T}\bullet\prod(\mathscr{C})\odot
\mathcal {X}_{X}.\Box$

\begin{example}\cite{Wang1}
Let $U=\{a,b,c,d,e,f\}$, and $\mathscr{C}=\{K_{1},K_{2},K_{3},K_{4}\}$, where $K_{1}=\{a,b\},K_{2}=\{a,c,d\},K_{3}=\{a,b,c,d\}$, and $K_{4}=\{d,e,f\}$. Then we have Tables 1 and 2(Tables 5 and 6 in Shiping Wang[Information Sciences 263(1),
186-197, 2014]) as follows:
\begin{table}[H]\renewcommand{\arraystretch}{1.2}
\caption{The sixth upper approximations computed by characteristic matrices. } \tabcolsep0.34in
\begin{tabular}{cccc}\hline
X&$\mathcal {X}_{X}$ & $(\prod(\mathscr{C}))^{T}\bullet\prod(\mathscr{C})\bullet
\mathcal {X}_{X}$& $XH_{\mathscr{C}}(X)$ \\\hline
$\{a\}$ &$[1 ~0 ~0 ~0~ 0 ~0]^{T}$    & $[1~ 1~ 1~ 1 ~0 ~0]^{T}$   &  $\{a,b,c,d\}$   \\
$\{a,b\}$ &$[1 ~1 ~0 ~0 ~0 ~0]^{T}$    & $[1~ 1 ~1 ~1~ 0 ~0]^{T}$   &  $\{a,b,c,d\}$   \\
$\{a,b,c\}$ &$[1 ~1~ 1~ 0~ 0 ~0]^{T}$    & $[1 ~1~ 1 ~1~ 0 ~0]^{T}$   &  $\{a,b,c,d\}$   \\
$\{d,e,f\}$ &$[0 ~0 ~0 ~1~ 1 ~1]^{T}$    & $[1 ~0~ 1~ 1 ~1 ~1]^{T}$   &  $\{a,c,d,e,f\}$   \\
$\{a,d,e,f\}$ &$[1 ~0 ~0~ 1 ~1 ~1]^{T}$    & $[1~ 1~ 1~ 1~ 1 ~1]^{T}$   & $\{a,b,c,d,e,f\}$
\\\hline
\end{tabular}
\end{table}

\begin{table}[H]\renewcommand{\arraystretch}{1.2}
\caption{The sixth lower approximations computed by characteristic matrices. } \tabcolsep0.3in
\begin{tabular}{cccc}\hline
X&$\mathcal {X}_{X}$ & $(\prod(\mathscr{C}))^{T}\bullet\prod(\mathscr{C})\odot
\mathcal {X}_{X}$& $XL_{\mathscr{C}}(X)$ \\\hline
$\{a\}$ &$[1 ~0 ~0 ~0 ~0 ~0]^{T}$    & $[0~ 0 ~0 ~0~ 0 ~0]^{T}$   &  $\emptyset$   \\
$\{a,b\}$ &$[1 ~1 ~0~ 0~ 0 ~0]^{T}$    & $[0 ~1 ~0 ~0 ~0~ 0]^{T}$   &  $\{b\}$   \\
$\{a,b,c\}$ &$[1 ~1 ~1 ~0~ 0 ~0]^{T}$    & $[0~ 1~ 0~ 0 ~0~ 0]^{T}$   &  $\{b\}$   \\
$\{a,b,c,d\}$ &$[1 ~1 ~1 ~1 ~0 ~0]^{T}$    & $[1 ~1 ~1 ~0 ~0 ~0]^{T}$   &  $\{a,b,c\}$   \\
$\{a,b,d,e,f\}$ &$[1 ~1 ~0 ~1~ 1~1]^{T}$    & $[0 ~1~ 0 ~0~ 1 ~1]^{T}$   &  $\{b,e,f\}$   \\
$\{a,b,c,d,e,f\}$ &$[1 ~1 ~1 ~1 ~1 ~1]^{T}$    & $[1 ~1~ 1 ~1 ~1 ~1]^{T}$   & $\{a,b,c,d,e,f\}$
\\\hline
\end{tabular}
\end{table}
\end{example}

\begin{example}
(Continued from Example 3.2) By Definition 2.1(3), we have Tables 3 and 4 as follows:
\begin{table}[H]\renewcommand{\arraystretch}{1.2}
\caption{The sixth upper approximations computed by characteristic matrices. } \tabcolsep1.24in
\begin{tabular}{cc}\hline
X& $XH_{\mathscr{C}}(X)$ \\\hline
$\{a\}$  &  $\{a,b,c,d\}$   \\
$\{a,b\}$  &  $\{a,b,c,d\}$   \\
$\{a,b,c\}$ &  $\{a,b,c,d\}$   \\
$\{d,e,f\}$ &  $\{a,c,d,e,f\}$   \\
$\{a,d,e,f\}$ & $\{a,b,c,d,e,f\}$
\\\hline
\end{tabular}
\end{table}

\begin{table}[H]\renewcommand{\arraystretch}{1.2}
\caption{The sixth lower approximations computed by characteristic matrices. } \tabcolsep1.15in
\begin{tabular}{cc}\hline
X&$XL_{\mathscr{C}}(X)$ \\\hline
$\{a\}$ & $\{a\}$   \\
$\{a,b\}$ &$\{a,b\}$   \\
$\{a,b,c\}$ &$\{a,b\}$   \\
$\{a,b,c,d\}$ &$\{a,b,c,d\}$   \\
$\{a,b,d,e,f\}$ & $\{a,b,d,e,f\}$   \\
$\{a,b,c,d,e,f\}$ &$\{a,b,c,d,e,f\}$
\\\hline
\end{tabular}
\end{table}
\end{example}

In Examples 3.2 and 3.3, we see the sixth lower approximations of sets in Table 2 are different from the results in Table 4.

\begin{theorem}
Let $(U,\mathscr{C})$ be a covering approximation space, where $U=\{x_{1},x_{2},...,x_{n}\}$,
$\mathscr{C}=\{C_{1},C_{2},...,C_{m}\}$, and $N(x)=\bigcap\{C_{i}|x\in C_{i}\in \mathscr{C}\}$ for any $x\in U$.
Then we have
$IL_{\mathscr{C}}(X)=XL_{\mathscr{C}}(X)$ for any $X\subseteq U$.
\end{theorem}
{\bf{Proof:}} For any $y\in IL_{\mathscr{C}}(X)=\{x|N(x)\subseteq X\}$, we get $N(y)\subseteq X$. Thus, $y\in XL_{\mathscr{C}}(X)=\bigcup\{N(x)|N(x)\subseteq X\}.$
For $z\in XL_{\mathscr{C}}(X)=\bigcup\{N(x)|N(x)\subseteq X\}$, there exists $x$ such that $z\in N(x)\subseteq X$. It follows that $N(z)\subseteq N(x)\subseteq X$. Thus, $z\in IL_{\mathscr{C}}(X)=\{x|N(x)\subseteq X\}$. Therefore, we have $
IL_{\mathscr{C}}(X)=XL_{\mathscr{C}}(X)$ for any $X\subseteq U.\Box$

By Theorem 3.4, we show the improvement for Theorem 7 in Shiping Wang[Information Sciences 263(1),
186-197, 2014] as follows.

\begin{theorem}
Let $(U,\mathscr{C})$ be a covering approximation space, where $U=\{x_{1},x_{2},...,x_{n}\}$,
$\mathscr{C}=\{C_{1},C_{2},...,C_{m}\}$, and
$\mathcal {X}_{X}$ the characteristic function of $X\subseteq U$. Then
\begin{eqnarray*}
\mathcal {X}_{XH_{\mathscr{C}}(X)}=(\prod(\mathscr{C}))^{T}\bullet\prod(\mathscr{C})\bullet
\mathcal {X}_{X}, \mathcal {X}_{XL_{\mathscr{C}}(X)}=\prod(\mathscr{C})\odot
\mathcal {X}_{X}.
\end{eqnarray*}
\end{theorem}
{\bf{Proof:}} We construct a covering
$Cov(\mathscr{C})=\{N(x)|x\in U\}$ represented by the matrix $(\prod(\mathscr{C}))^{T}$. By Theorem 2.3(1), we obtain $XH_{\mathscr{C}}(X)=SH_{Cov(\mathscr{C})}(X)$ for any $X\subseteq U$. Furthermore, by Theorem 3.4, we get $XL_{\mathscr{C}}(X)=IL_{\mathscr{C}}(X)$ for any $X\subseteq U$. By Theorem 2.3(2), we have $\mathcal {X}_{IL_{\mathscr{C}}(X)}=\prod(\mathscr{C})\odot
\mathcal {X}_{X}$ for any $X\subseteq U$. Therefore, $\mathcal {X}_{XH_{\mathscr{C}}(X)}=(\prod(\mathscr{C}))^{T}\bullet\prod(\mathscr{C})\bullet
\mathcal {X}_{X}$ and $\mathcal {X}_{XL_{\mathscr{C}}(X)}=\prod(\mathscr{C})\odot
\mathcal {X}_{X}.\Box$

\begin{example}(Continued from Example 3.2) By Theorem 3.5, we have Tables 5 and 6 as follows:
\begin{table}[H]\renewcommand{\arraystretch}{1.2}
\caption{The sixth upper approximations computed by characteristic matrices. } \tabcolsep0.335in
\begin{tabular}{cccc}\hline
X&$\mathcal {X}_{X}$ & $(\prod(\mathscr{C}))^{T}\bullet\prod(\mathscr{C})\bullet
\mathcal {X}_{X}$& $XH_{\mathscr{C}}(X)$ \\\hline
$\{a\}$ &$[1 ~0 ~0 ~0~ 0 ~0]^{T}$    & $[1 ~1 ~1 ~1~ 0 ~0]^{T}$   &  $\{a,b,c,d\}$   \\
$\{a,b\}$ &$[1 ~1~ 0 ~0 ~0 ~0]^{T}$    & $[1 ~1 ~1 ~1 ~0 ~0]^{T}$   &  $\{a,b,c,d\}$   \\
$\{a,b,c\}$ &$[1 ~1 ~1 ~0 ~0~ 0]^{T}$    & $[1 ~1~ 1~ 1 ~0~ 0]^{T}$   &  $\{a,b,c,d\}$   \\
$\{d,e,f\}$ &$[0 ~0 ~0 ~1 ~1 ~1]^{T}$    & $[1 ~0~ 1 ~1 ~1 ~1]^{T}$   &  $\{a,c,d,e,f\}$   \\
$\{a,d,e,f\}$ &$[1 ~0 ~0 ~1 ~1 ~1]^{T}$    & $[1 ~1 ~1 ~1~ 1 ~1]^{T}$   & $\{a,b,c,d,e,f\}$
\\\hline
\end{tabular}
\end{table}

\begin{table}[H]\renewcommand{\arraystretch}{1.2}
\caption{The sixth lower approximations computed by characteristic matrices. } \tabcolsep0.37in
\begin{tabular}{cccc}\hline
X&$\mathcal {X}_{X}$ & $\prod(\mathscr{C})\odot
\mathcal {X}_{X}$& $XL_{\mathscr{C}}(X)$ \\\hline
$\{a\}$ &$[1 ~0 ~0 ~0 ~0 ~0]^{T}$    & $[1 ~0 ~0 ~0 ~0 ~0]^{T}$   & $\{a\}$   \\
$\{a,b\}$ &$[1 ~1 ~0 ~0 ~0 ~0]^{T}$    & $[1 ~1 ~0 ~0 ~0 ~0]^{T}$   &  $\{a,b\}$   \\
$\{a,b,c\}$ &$[1 ~1 ~1 ~0 ~0 ~0]^{T}$    & $[1 ~1 ~0 ~0 ~0 ~0]^{T}$   &  $\{a,b\}$   \\
$\{a,b,c,d\}$ &$[1 ~1 ~1 ~1 ~0 ~0]^{T}$    & $[1 ~1 ~1 ~1 ~0 ~0]^{T}$   &  $\{a,b,c,d\}$   \\
$\{a,b,d,e,f\}$ &$[1 ~1 ~0 ~1 ~1 ~1]^{T}$    & $[1 ~1 ~0 ~1 ~1 ~1]^{T}$   &  $\{a,b,d,e,f\}$   \\
$\{a,b,c,d,e,f\}$ &$[1 ~1 ~1 ~1 ~1 ~1]^{T}$    & $[1 ~1 ~1 ~1 ~1 ~1]^{T}$   & $\{a,b,c,d,e,f\}$
\\\hline
\end{tabular}
\end{table}
\end{example}

\begin{theorem}
Let $(U,\mathscr{C})$ be a covering approximation space, where $U=\{x_{1},x_{2},...,x_{n}\}$,
$\mathscr{C}=\{C_{1},C_{2},...,C_{m}\}$, $Cov(\mathscr{C})=\{N(x)|x\in U\}$ induced by $\mathscr{C}$, and $N^{\ast}(x)=\bigcap\{N(x_{i})|x\in N(x_{i})\in Cov(\mathscr{C})\}$. Then we have $IL_{\mathscr{C}}(X)=IL_{Cov(\mathscr{C})}(X)$ for any $X\subseteq U$.
\end{theorem}
{\bf{Proof:}} By Definition 2.1, we have $IL_{\mathscr{C}}(X)=\{x|N(x)\subseteq X\}$ and $IL_{Cov(\mathscr{C})}(X)=\{x|N^{\ast}(x)\subseteq X\}$ for any $X\subseteq U$.
For any $x\in IL_{\mathscr{C}}(X)$, there exists $x\in U$ such that $N(x)\subseteq X$. Since $
N^{\ast}(x)=\bigcap\{N(x_{i})|x\in N(x_{i})\in Cov(\mathscr{C})\}$, we obtain $N^{\ast}(x)\subseteq N(x)\subseteq X$.
It implies $x\in IL_{Cov(\mathscr{C})}(X)$.
For any $y\in IL_{Cov(\mathscr{C})}(X)=\{x|N^{\ast}(x)\subseteq X\}$, we get $N^{\ast}(y)\subseteq X$.
Since $N^{\ast}(y)=\bigcap\{N(x_{k_{i}})|y\in N(x_{k_{i}})\in Cov(\mathscr{C})\}$, we get $N^{\ast}(y)=N(y)\cap(\bigcap\{N(x_{k_{i}})|y\in N(x_{k_{i}})\in Cov(\mathscr{C}),x_{k_{i}}\neq y\})$.
If $y\in N(x_{k_{i}})$, then $N(y)\subseteq N(x_{k_{i}})$. It follows that
$N^{\ast}(y)=N(y)$. Thus, $y\in IL_{\mathscr{C}}(X)$. Therefore, we obtain $IL_{\mathscr{C}}(X)=IL_{Cov(\mathscr{C})}(X)$ for any $X\subseteq U.\Box$

\begin{proposition}
Let $(U,\mathscr{C})$ be a covering approximation space, where $U=\{x_{1},x_{2},...,x_{n}\}$, and
$\mathscr{C}=\{C_{1},C_{2},...,C_{m}\}$. Then
$XL_{\mathscr{C}}(X)=XL_{Cov(\mathscr{C})}(X).
$
\end{proposition}
{\bf{Proof:}} By Theorems 3.4 and 3.7, the proof is straightforward.$\Box$

\begin{proposition}
Let $(U,\mathscr{C})$ be a covering approximation space, where $U=\{x_{1},x_{2},...,x_{n}\}$,
$\mathscr{C}=\{C_{1},C_{2},...,C_{m}\}$, and $\mathcal {X}_{X}$ the characteristic function of $X\subseteq U$. Then
\begin{eqnarray*}
\mathcal {X}_{XL_{\mathscr{C}}(X)}=(\prod(\mathscr{C}))^{T}\odot\prod(\mathscr{C})\odot
\mathcal {X}_{X}.
\end{eqnarray*}
\end{proposition}
{\bf{Proof:}} By Theorems 2.3, 3.5, and 3.7, and Proposition 3.8, the proof is straightforward.$\Box$

\begin{definition}
Let $(U,\mathscr{C})$ be a covering approximation space, where $U=\{x_{1},x_{2},...,x_{n}\}$,
$\mathscr{C}=\{C_{1},C_{2},...,C_{m}\}$, and $N(x)=\bigcap\{C_{i}|x\in C_{i}\in \mathscr{C}\}$ for any $x\in U$. For any
$X\subseteq U$, the sixth dual upper and lower approximations
of $X$ with respect to $\mathscr{C}$ are defined as follows:
\begin{eqnarray*}
XH^{d}_{\mathscr{C}}(X)=\bigcup\{N(x)|N(x)\cap X\neq \emptyset\},
XL^{d}_{\mathscr{C}}(X)=[XH^{d}_{\mathscr{C}}(X^{c})]^{c}.
\end{eqnarray*}
\end{definition}

The following theorem illustrates how to construct the sixth dual upper and lower approximations
of sets from the view of matrix.

\begin{theorem}
Let $(U,\mathscr{C})$ be a covering approximation space, where $U=\{x_{1},x_{2},...,x_{n}\}$,
$\mathscr{C}=\{C_{1},C_{2},...,C_{m}\}$, and
$\mathcal {X}_{X}$ the characteristic function of $X\subseteq U$. Then
\begin{eqnarray*}
\mathcal {X}_{XH^{d}_{\mathscr{C}}(X)}=(\prod(\mathscr{C}))^{T}\bullet\prod(\mathscr{C})\bullet
\mathcal {X}_{X}, \mathcal {X}_{XL^{d}_{\mathscr{C}}(X)}=(\prod(\mathscr{C}))^{T}\bullet\prod(\mathscr{C})\odot
\mathcal {X}_{X}.
\end{eqnarray*}
\end{theorem}
{\bf{Proof:}} We construct a covering
$Cov(\mathscr{C})=\{N(x)|x\in U\}$ represented by the matrix $(\prod(\mathscr{C}))^{T}$. By Definition 2.1, we obtain $XH^{d}_{\mathscr{C}}(X)=SH_{Cov(\mathscr{C})}(X)$ and $XL^{d}_{\mathscr{C}}(X)=SL_{Cov(\mathscr{C})}(X)$ for any $X\subseteq U$. By Theorem 2.3, we have $\mathcal {X}_{XH^{d}_{\mathscr{C}}(X)}=(\prod(\mathscr{C}))^{T}\bullet\prod(\mathscr{C})\bullet
\mathcal {X}_{X}$ and $\mathcal {X}_{XL^{d}_{\mathscr{C}}(X)}=(\prod(\mathscr{C}))^{T}\bullet\prod(\mathscr{C})\odot
\mathcal {X}_{X}.\Box$

\begin{example}(Continued from Example 3.6) By Definition 3.10 and Theorem 3.11, we have Tables 7 and 8 as follows:
\begin{table}[H]\renewcommand{\arraystretch}{1.2}
\caption{The sixth upper approximations computed by characteristic matrices. } \tabcolsep0.34in
\begin{tabular}{cccc}\hline
X&$\mathcal {X}_{X}$ & $(\prod(\mathscr{C}))^{T}\bullet\prod(\mathscr{C})\bullet
\mathcal {X}_{X}$& $XH^{d}_{\mathscr{C}}(X)$ \\\hline
$\{a\}$ &$[1 ~0 ~0 ~0~ 0 ~0]^{T}$    & $[1~ 1~ 1~ 1 ~0 ~0]^{T}$   &  $\{a,b,c,d\}$   \\
$\{a,b\}$ &$[1 ~1 ~0 ~0 ~0 ~0]^{T}$    & $[1~ 1 ~1 ~1~ 0 ~0]^{T}$   &  $\{a,b,c,d\}$   \\
$\{a,b,c\}$ &$[1 ~1~ 1~ 0~ 0 ~0]^{T}$    & $[1 ~1~ 1 ~1~ 0 ~0]^{T}$   &  $\{a,b,c,d\}$   \\
$\{d,e,f\}$ &$[0 ~0 ~0 ~1~ 1 ~1]^{T}$    & $[1 ~0~ 1~ 1 ~1 ~1]^{T}$   &  $\{a,c,d,e,f\}$   \\
$\{a,d,e,f\}$ &$[1 ~0 ~0~ 1 ~1 ~1]^{T}$    & $[1~ 1~ 1~ 1~ 1 ~1]^{T}$   & $\{a,b,c,d,e,f\}$
\\\hline
\end{tabular}
\end{table}

\begin{table}[H]\renewcommand{\arraystretch}{1.2}
\caption{The sixth lower approximations computed by characteristic matrices. } \tabcolsep0.3in
\begin{tabular}{cccc}\hline
X&$\mathcal {X}_{X}$ & $(\prod(\mathscr{C}))^{T}\bullet\prod(\mathscr{C})\odot
\mathcal {X}_{X}$& $XL^{d}_{\mathscr{C}}(X)$ \\\hline
$\{a\}$ &$[1 ~0 ~0 ~0 ~0 ~0]^{T}$    & $[0~ 0 ~0 ~0~ 0 ~0]^{T}$   &  $\emptyset$   \\
$\{a,b\}$ &$[1 ~1 ~0~ 0~ 0 ~0]^{T}$    & $[0 ~1 ~0 ~0 ~0~ 0]^{T}$   &  $\{b\}$   \\
$\{a,b,c\}$ &$[1 ~1 ~1 ~0~ 0 ~0]^{T}$    & $[0~ 1~ 0~ 0 ~0~ 0]^{T}$   &  $\{b\}$   \\
$\{a,b,c,d\}$ &$[1 ~1 ~1 ~1 ~0 ~0]^{T}$    & $[1 ~1 ~1 ~0 ~0 ~0]^{T}$   &  $\{a,b,c\}$   \\
$\{a,b,d,e,f\}$ &$[1 ~1 ~0 ~1~ 1~1]^{T}$    & $[0 ~1~ 0 ~0~ 1 ~1]^{T}$   &  $\{b,e,f\}$   \\
$\{a,b,c,d,e,f\}$ &$[1 ~1 ~1 ~1 ~1 ~1]^{T}$    & $[1 ~1~ 1 ~1 ~1 ~1]^{T}$   & $\{a,b,c,d,e,f\}$
\\\hline
\end{tabular}
\end{table}
\end{example}

The following example is employed to illustrate there are some differences between the sixth lower approximation operator and the sixth dual lower approximation operator.

\begin{example}(Continued from Examples 3.6 and 3.12)
 For $X=\{a,b,c\}$, by Definition 2.1, we have $XL_{\mathscr{C}}(X)=\{a,b\}$. Furthermore, for $X^{c}=\{d,e,f\}$, we get $XL^{d}_{\mathscr{C}}(X)=[XH_{\mathscr{C}}(X^{c})]^{c}=\{b\}$. Therefore, we obtain $XL_{\mathscr{C}}(X)\neq XL^{d}_{\mathscr{C}}(X)$.
\end{example}

\section{Conclusions}

In this paper, we have provided some improvements for	
Theorem 7 and Example 8 in Shiping Wang [Information Sciences 263(1),
186-197, 2014]. Furthermore, we have presented the sixth dual lower and upper approximations of sets for covering approximation spaces. We have constructed the sixth dual lower and upper approximations of sets from the view of matrix.
In the future, we will further study the lower and upper approximation operators for covering approximation spaces.

\section*{ Acknowledgments}

We would like to thank the anonymous reviewers very much for their
professional comments and valuable suggestions. This work is
supported by the National Natural Science Foundation of China (NO.
61273304,11401052,11526039), Doctoral Fund of Ministry of Education of China(No.201300721004), China Postdoctoral Science Foundation(NO.2015M580353), the Scientific
Research Fund of Hunan Provincial Education Department(No.14C0049,15B004).

\end{document}